\documentclass{emulateapj}

\def\ts     {\thinspace}
\def\kms    {\ts km\ts s$^{-1}$}
\def\etal   {{\rm et\ts al.}}
\def\msol   {$M_{\odot}$}
\def\lsol   {$L_{\odot}$}

\def\aco    {{\rm CO}($J$=1$\to$0)}

\def\cco    {{\rm CO}($J$=3$\to$2)}

\def\fco    {{\rm CO}($J$=6$\to$5)}
\def\gco    {{\rm CO}($J$=7$\to$6)}

\def\ahcn    {{\rm HCN}($J$=1$\to$0)}
\def\bhcn    {{\rm HCN}($J$=2$\to$1)}

\def\bhco    {{\rm HCO$^+$}($J$=2$\to$1)}

\def\bhnc    {{\rm HNC}($J$=2$\to$1)}
\def\bcch    {{\rm C$_2$H}($N$=2$\to$1)}

\slugcomment{draft version \today, accepted for publication in the Astrophysical Journal Letters}

\shorttitle{Search for HCN(2-1) at $z$=6.42}
\shortauthors{Riechers et al.}

\begin{document}

\title{Observations of Dense Molecular Gas in a Quasar Host Galaxy at
  $z$=6.42: 
  Further Evidence for a Non-Linear Dense Gas -- Star Formation
  Relation at Early Cosmic Times}

\author{Dominik A. Riechers\altaffilmark{1}, Fabian
  Walter\altaffilmark{1}, Christopher L. Carilli\altaffilmark{2}, and
  Frank Bertoldi\altaffilmark{3} }

\altaffiltext{1}{Max-Planck-Institut f\"ur Astronomie, K\"onigstuhl 17, 
Heidelberg, D-69117, Germany; riechers@mpia.de}

\altaffiltext{2}{National Radio Astronomy Observatory, PO Box O,
  Socorro, NM 87801, USA}

\altaffiltext{3}{Argelander-Institut f\"ur Astronomie, Universit\"at
  Bonn, Auf dem H\"ugel 71, Bonn, D-53121, Germany}


\begin{abstract}
  We report a sensitive search for the \bhcn\ emission line towards
  SDSS\,J114816.64+525150.3 (hereafter:\ J1148+5251) at $z$=6.42 with
  the Very Large Array (VLA).  HCN emission is a star formation
  indicator, tracing dense molecular hydrogen gas ($n({\rm H_2}) \geq
  10^4\,$cm$^{-3}$) within star-forming molecular clouds.  No emission
  was detected in the deep interferometer maps of J1148+5251.  We
  derive a limit for the HCN line luminosity of $L'_{\rm HCN} < 3.3
  \times 10^{9}\,$K \kms pc$^2$, corresponding to a HCN/CO luminosity
  ratio of $L'_{\rm HCN}$/$L'_{\rm CO}$$<$0.13. This limit is
  consistent with a fraction of dense molecular gas in J1148+5251
  within the range of nearby ultraluminous infrared galaxies (ULIRGs;
  median value:\ $L'_{\rm HCN}$/$L'_{\rm CO}$=0.17$^{+0.05}_{-0.08}$)
  and HCN-detected $z$$>$2 galaxies (0.17$^{+0.09}_{-0.08}$).  The
  relationship between $L'_{\rm HCN}$ and $L_{\rm FIR}$ is considered
  to be a measure for the efficiency at which stars form out of dense
  gas.  In the nearby universe, these quantities show a linear
  correlation, and thus, a practically constant average ratio.  In
  J1148+5251, we find $L_{\rm FIR}$/$L'_{\rm HCN}$$>$6600. This is
  significantly higher than the average ratios for normal nearby
  spiral galaxies ($L_{\rm FIR}$/$L'_{\rm HCN}$=580$^{+510}_{-270}$)
  and ULIRGs (740$^{+505}_{-50}$), but consistent with a rising trend
  as indicated by other $z$$>$2 galaxies (predominantly quasars;
  1525$^{+1300}_{-475}$).  It is unlikely that this rising trend can
  be accounted for by a contribution of active galactic nucleus (AGN)
  heating to $L_{\rm FIR}$ alone, and may hint at a higher median gas
  density and/or elevated star-formation efficiency toward the more
  luminous high-redshift systems.  There is marginal evidence that the
  $L_{\rm FIR}$/$L'_{\rm HCN}$ ratio in J1148+5251 may even exceed the
  rising trend set by other $z$$>$2 galaxies; however, only future
  facilities with very large collecting areas such as the Square
  Kilometre Array (SKA) will offer the sensitivity required to further
  investigate this question.
\end{abstract}


\section{Introduction}

High redshift galaxy populations are now being detected back to 780
million years after the Big Bang (spectroscopically confirmed:\
$z$=6.96; Iye \etal\ \citeyear{iye06}), probing into the epoch of
cosmic reionization (e.g., Fan \etal\ \citeyear{fan06}; Hu \& Cowie
\citeyear{hu06}). Many of these very distant galaxies show evidence
for star formation activity (e.g., Taniguchi \etal\ \citeyear{tan05}).
Some are even found to be hyperluminous infrared galaxies (HLIRGs;
Bertoldi \etal\ \citeyear{ber03a}; Wang \etal\ \citeyear{wan07}) with
far-infrared (FIR) luminosities exceeding 10$^{13}$\,\lsol, suggesting
vigorous star formation and/or AGN activity.  To probe the earliest
stages of galaxy formation and the importance of AGN in this process,
it is necessary to study the star formation characteristics of these
galaxies.

A good diagnostic to examine the star-forming environments in distant
HLIRGs are observations of molecular gas, the fuel for star formation.
The by far brightest and most common indicator of molecular gas in
galaxies is line emission from the rotational transitions of carbon
monoxide (CO), which was detected in $\sim$40 galaxies at high
redshift ($z$$>$1; see Solomon \& Vanden Bout \citeyear{sv05} for a
review). These observations have revealed molecular gas reservoirs
with masses of $>$10$^{10}$\,\msol\ in these galaxies, even in the
highest redshift quasar known, J1148+5251 at $z$=6.42 (Walter \etal\
\citeyear{wal03}, \citeyear{wal04}; Bertoldi \etal\
\citeyear{ber03b}).

Although CO is a good tracer of the total amount of molecular gas in a
galaxy, due to the relatively low critical density of $n_{\rm H_2}
\sim 10^2-10^3\,$cm$^{-3}$ required to collisionally excite its lower
$J$ transitions, it is not a reliable tracer of the dense molecular
cloud cores where the actual star formation takes place.  Recent
studies of nearby actively star-forming galaxies have shown that
hydrogen cyanide (HCN) is a far better tracer of the dense ($n_{\rm
  H_2} \sim 10^5-10^6$\,cm$^{-3}$) molecular gas where stars actually
form (e.g.\ Gao \& Solomon \citeyear{gao04a}, \citeyear{gao04b},
hereafter:\ GS04a, GS04b).  In the local universe it was found that
the HCN luminosity ($L'_{\rm HCN}$) scales linearly (unlike $L'_{\rm
  CO}$) with the FIR luminosity ($L_{\rm FIR}$) over 7--8 orders of
magnitude, ranging from Galactic dense cores to ULIRGs (Wu et al.\
\citeyear{wu05}).  As $L_{\rm FIR}$ traces the massive star formation
rate (unless AGN heating is significant), this implies that HCN is
also a good tracer of star formation.

HCN has now also been detected in five galaxies at $z$$>$2 (Solomon
\etal\ \citeyear{sol03}; Vanden Bout \etal\ \citeyear{vdb04}; Carilli
\etal\ \citeyear{car05}, hereafter:\ C05; Wagg \etal\
\citeyear{wag05}; Gao \etal\ \citeyear{gao07}, hereafter:\ G07).
Adding a number of upper limits obtained for other high-$z$ galaxies,
these observations indicate that the more luminous, higher redshift
systems systematically deviate from the linear $L'_{\rm HCN}$--$L_{\rm
  FIR}$ correlation found in the local universe (G07), and hint at a
rising slope of the relation toward high $L_{\rm FIR}$ and/or $z$.  To
further investigate this apparent non-linear, rising trend, our aim
has been to extend the range of existing HCN observations beyond
redshift 6 and to higher $L_{\rm FIR}$.

In this letter, we report sensitive VLA\footnote{The Very Large Array
  is a facility of the National Radio Astronomy Observatory, operated
  by Associated Universities, Inc., under cooperative agreement with
  the National Science Foundation.} observations of \bhcn\ emission
toward the $z$=6.42 quasar J1148+5251, the highest redshift source
detected in CO.  A previous, less sensitive search for \bhcn\ emission
in this source has yielded no detection (C05).  We use a concordance,
flat $\Lambda$CDM cosmology throughout, with
$H_0$=71\,\kms\,Mpc$^{-1}$, $\Omega_{\rm M}$=0.27, and
$\Omega_{\Lambda}$=0.73 (Spergel \etal\ \citeyear{spe06}).

\section{Observations}

\begin{figure}
\epsscale{1.2}
\plotone{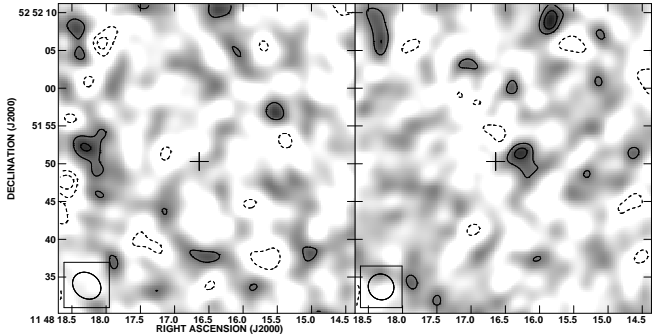}
\vspace{-5mm}

\caption{VLA observations of \bhcn\ ({\em left}) and 23.9\,GHz
continuum emission ({\em right}) towards J1148+5251 at a
resolution of 4.0\,$''$$\times$3.3\,$''$ ({\em left}) and
3.5\,$''$$\times$3.3\,$''$ ({\em right}; as indicated in the bottom
left corner of both panels). The \bhcn\ line map is integrated over the
full 25\,MHz IF.  No emission is detected in both maps.  The cross
indicates the geometrical center of the CO emission.  Contours are
shown at (-3, -2, 2, 3)$\times\sigma$ (1$\sigma = 15\,\mu$Jy
beam$^{-1}$ for the line map, and 14\,$\mu$Jy beam$^{-1}$ for the
continuum map).
\label{f1}}
\end{figure}

We observed the \bhcn\ transition line ($\nu_{\rm rest} =
177.2612230\,$GHz) toward J1148+5251 using the VLA in D configuration
in 10\,tracks on 2004 June 21 and 26 (these observations were
discussed by C05), and between 2007 May 01 and 13. At the target
redshift of 6.419, the line is shifted to 23.892873\,GHz (11.96\,mm).
The total integration time amounts to 80\,hr.  Observations were
performed in fast-switching mode using the nearby source 11534+49311
(at $3.4^\circ$ distance to J1148+5251) for secondary amplitude and
phase calibration.  Observations were carried out under very good
weather conditions with 26 antennas [including 9 Expanded Very Large
Array (EVLA) antennas in 2007]:\ the phase stability was excellent
(typically $<$15$^\circ$ phase rms for the longest baselines).  The
QSO 3C286 was observed as a primary flux calibrator.  The resulting
flux scale is accurate within the standard 15\% uncertainty.

In 2004, two 25\,MHz wide intermediate frequency bands (IFs) with
seven 3.125\,MHz channels each were observed simultaneously, one
centered at the \bhcn\ line frequency (23.8929\,GHz), and one centered
at 24.0430\,GHz for continuum monitoring\footnote{In fact, this is the
  center frequency of the \bhco\ line, which however was not detected.
  Neither was the continuum.}. This leads to an effective bandwidth of
21.875\,MHz for both the line and continuum observations,
corresponding to 274\,\kms\ at 23.9\,GHz. For lines as broad as CO
(279\,\kms\ FWHM; Bertoldi \etal\ \citeyear{ber03b}), this setup would
miss only $\lesssim$15\% flux in the line wings.  In 2007,
observations were carried out in continuum mode, with one 25\,MHz wide
IF centered at the \bhcn\ line frequency (23.8929\,GHz), and one
50\,MHz wide IF centered at 23.7649\,GHz to monitor the source's
continuum\footnote{This frequency avoids other potentially bright
  emission lines near the \bhcn\ frequency, such as \bhco, \bhnc, and
  \bcch.}.

For data reduction and analysis, the ${\mathcal{AIPS}}$ package was
used. All data were mapped using the CLEAN algorithm and `natural'
weighting; this results in synthesized beams of
4.0\,$''$$\times$3.3\,$''$ for the line map and
3.5\,$''$$\times$3.3\,$''$ ($\sim$20\,kpc at $z = 6.42$) for the
continuum map (see Fig.~\ref{f1}). The final rms in the integrated
\bhcn\ line map is 15\,$\mu$Jy beam$^{-1}$, and 14\,$\mu$Jy
beam$^{-1}$ in the continuum map.  Averaging both uv datasets to a
`high-sensitivity' continuum map leads to an rms of 11\,$\mu$Jy
beam$^{-1}$.

\begin{deluxetable}{ r c c c }
\tabletypesize{\scriptsize}
\tablecaption{Line luminosities in SDSS\,J1148+5251. \label{tab-1}}
\tablehead{
& $S_{\nu}$ & $L'$ & Ref. \\
& [$\mu$Jy] & [10$^9$\,K\,\kms\,pc$^2$] & }
\startdata
\bhcn\ & (7 $\pm$ 15)   & $<$3.3         & 1 \\
\aco\  & $<$360         & $<$142         & 2 \\ 
\cco\  & 570 $\pm$ 57   & 26.4 $\pm$ 2.6 & 3 \\ 
\fco\  & 2450           & 26.9 $\pm$ 2.4 & 2 \\ 
\gco\  & 2140           & 17.3 $\pm$ 2.4 & 2 \\ 
\vspace{-2mm}
\enddata 
\tablerefs{${}$[1] This work, [2] Bertoldi \etal\ (\citeyear{ber03b}), 
[3] Walter \etal\ (\citeyear{wal03}).
}
\tablecomments{${}$ Bracketed number indicates nondetection.}
\end{deluxetable}


\section{Results}

No \bhcn\ emission is detected in J1148+5251 (Fig.~\ref{f1}, left). We
derive a peak flux density of (7 $\pm$ 15)\,$\mu$Jy beam$^{-1}$ at the
source's CO position (see Table~\ref{tab-1}), setting an upper
limit\footnote{In this letter, we quote 2$\sigma$ limits; however,
  note that all conclusions hold for 3$\sigma$ limits.} of 30\,$\mu$Jy
to the emission line peak flux.  No continuum emission is detected at
and/or close to the \bhcn\ line frequency (Fig.~\ref{f1}, right). From
the `high-sensitivity' continuum map, we derive a peak flux density of
(11 $\pm$ 11)\,$\mu$Jy beam$^{-1}$ at the source's position, setting
an upper limit of 22\,$\mu$Jy to the 23.9\,GHz continuum flux density.
This is consistent with the model-predicted continuum level of
$\lesssim$8\,$\mu$Jy (Beelen \etal\ \citeyear{bee06}).

From our observations, we derive a limit to the \bhcn\ line luminosity
of $L'_{\rm HCN} < 3.3 \times 10^9\,$K\,\kms\,pc$^2$ (assuming a
HCN/CO linewidth ratio of 0.67, i.e., the average value for the four
$z$$>$2 HCN-detected quasars\footnote{This ratio lies between 57\% and
  80\% for the $z$$>$2 sample, but is up to 100\% in nearby starburst
  galaxies like NGC\,253 (e.g., Knudsen \etal\ \citeyear{knu07}).};
see Table~\ref{tab-2} for references), corresponding to 13\% of the CO
luminosity (Walter \etal\ \citeyear{wal03}).  Using the FIR luminosity
derived by Beelen \etal\ (\citeyear{bee06})\footnote{FIR luminosities
  estimated from modeling the sparsely sampled FIR SEDs of high-$z$
  galaxies are only accurate within a factor of 2.}, we find $L_{\rm
  FIR}$/$L'_{\rm HCN}$$>$6600.

\section{Analysis}

The nondetection of \bhcn\ in J1148+5251 at the depth of our
observations has several implications. For the following comparison
with other galaxies (for which $L'_{\rm CO}$ and $L'_{\rm HCN}$ are
given in the ground-state transitions), we assume that the \bhcn\ and
\cco\ lines are thermalized, so that $L'_{{\rm HCN}(J=2-1)}$=$L'_{{\rm
    HCN}(J=1-0)}$ and $L'_{{\rm CO}(J=3-2)}$=$L'_{{\rm CO}(J=1-0)}$.
Considering the high CO excitation in this source (Bertoldi \etal\
\citeyear{ber03b}) and the fact that both lines arise from low $J$
transitions, this is likely a valid assumption.

\begin{deluxetable}{ l c c l }
\tabletypesize{\scriptsize}
\tablecaption{Luminosity ratios at high $z$ and average values. \label{tab-2}}
\tablehead{
Source & $L_{\rm FIR}$/$L'_{\rm HCN}$ & $L'_{\rm HCN}$/$L'_{\rm CO}$ & Refs. \\
& [\lsol/$L_l$] & & }
\startdata
SDSS\,J1148+5251      & $>$6600 & $<$0.13 & 1,2,3 \\
\tableline
Cloverleaf            & 1305$\pm$360 & 0.10$\pm$0.02 & 4,5,6 \\
IRAS\,F10214+4724     & 2835$\pm$855 & 0.18$\pm$0.05 & 7,8,9 \\
VCV\,J1409+5628       & 2615$\pm$860 & 0.09$\pm$0.03 & 9,10,3 \\
APM\,08279+5255\tablenotemark{a} & 1000    & 0.27    & 11,12,13 \\
SMM\,J16359+6612      & 1550$\pm$550 & 0.18$\pm$0.04 & 14,15,16 \\
\tableline
\vspace{-2.5mm}
& & & \\
avg.\ `normal' spiral (33) & 580$^{+510}_{-270}$ & 0.04$^{+0.01}_{-0.02}$ & 17,14\\
avg.\ $z$$\simeq$0 LIRG (23) & 650$^{+360}_{-290}$ & 0.07$^{+0.02}_{-0.02}$ & 17,14\\
avg.\ $z$$\simeq$0 ULIRG (9) & 740$^{+505}_{-50}$ & 0.17$^{+0.05}_{-0.08}$ & 17,14\\
avg.\ $z$$\simeq$0 `all' (33+23+9) & 680$^{+400}_{-360}$ & 0.05$^{+0.05}_{-0.02}$ & 17,14\\
avg.\ $z$$>$2 galaxy (5) & 1525$^{+1300}_{-475}$ & 0.17$^{+0.09}_{-0.08}$ & 4--16\\
\vspace{-2mm}
\enddata 
\tablenotetext{a}{
We adopt the extrapolated \ahcn\ luminosity from Wei\ss\ \etal\
(\citeyear{wei07}), derived by assuming collisional
excitation only (uncertainties are model-dominated). 
Also, we only consider the `cold, dense' gas and dust component 
in their model, which is assumed to give rise to the HCN emission.}
\tablerefs{${}$[1] This work, 
[2] Walter \etal\ (\citeyear{wal03}), 
[3] Beelen \etal\ (\citeyear{bee06}), 
[4] Solomon \etal\ (\citeyear{sol03}), 
[5] Wei\ss\ \etal\ (\citeyear{wei03}), 
[6] Riechers \etal\ (\citeyear{rie06a}),
[7] Vanden Bout \etal\ (\citeyear{vdb04}),
[8] D.~Downes \& P.~M.~Solomon, in prep.,
[9] Carilli \etal\ (\citeyear{car05}),
[10] Beelen \etal\ (\citeyear{bee04}),
[11] Wagg \etal\ (\citeyear{wag05}),
[12] Riechers \etal\ (\citeyear{rie06b}),
[13] Wei\ss\ \etal\ (\citeyear{wei07}),
[14] Gao \etal\ (\citeyear{gao07}),
[15] Kneib \etal\ (\citeyear{kne05}),
[16] Kneib \etal\ (\citeyear{kne04}),
[17] Gao \& Solomon (\citeyear{gao04b}).}
\tablecomments{${}$ Line luminosity unit is $L_l$=K\,\kms\,pc$^2$. Bracketed numbers are sample sizes (`all' corresponds to the full Gao \& Solomon sample). For the averaged samples, median values and 1$\sigma$ statistical errors of the cumulative distribution are given. Where not quoted in the literature, 25\% error are assumed for $L_{\rm FIR}$.}
\end{deluxetable}


\subsection{`Dense Gas Fraction':\ The $L'_{\rm HCN}$/$L'_{\rm CO}$
  Ratio}

\aco\ emission is considered a good tracer for the total amount of
molecular gas in a galaxy, while \ahcn\ emission is considered a good
tracer for the dense peaks of the molecular mass distribution. The
$L'_{\rm HCN}$/$L'_{\rm CO}$ ratio thus is considered a measure for
the dense fraction of molecular gas in a galaxy\footnote{We note that
  Graci\'a-Carpio \etal\ (\citeyear{gra06}) have questioned the
  validity of $L'_{\rm HCN}$/$L'_{\rm CO}$ as a tracer of the dense
  gas fraction.}.  On average, this ratio is about four to five times
higher in local ULIRGs compared to normal, nearby spiral galaxies
[median value of 0.17$^{+0.05}_{-0.08}$ (ULIRGs)
vs.~0.04$^{+0.01}_{-0.02}$ (spirals); see Table~\ref{tab-2}].
However, there does not appear to be a further increase toward the
even more FIR-luminous systems at $z$$>$2 (0.17$^{+0.09}_{-0.08}$ on
average) within the statistical uncertainties (all values rederived
from GS04ab; G07).  The upper limit for $L'_{\rm HCN}$/$L'_{\rm CO}$
in J1148+5251 is 0.13, placing it below the median but within the
range of values measured for ULIRGs and $z$$>$2 galaxies
(Fig.~\ref{f2}a).  Our observations thus confirm the finding of G07
that galaxies brighter than $L_{\rm FIR}$=10$^{12}$\,\lsol\ appear to
have a higher fraction of dense molecular gas than normal spiral
galaxies; however, there is no indication that this fraction rises
further toward the highest $L_{\rm FIR}$, or that it changes with
redshift.

\subsection{`Star Formation Law':\ The $L_{\rm FIR}$/$L'_{\rm HCN}$
  Ratio}

\begin{figure*}
\epsscale{1.15}
\plotone{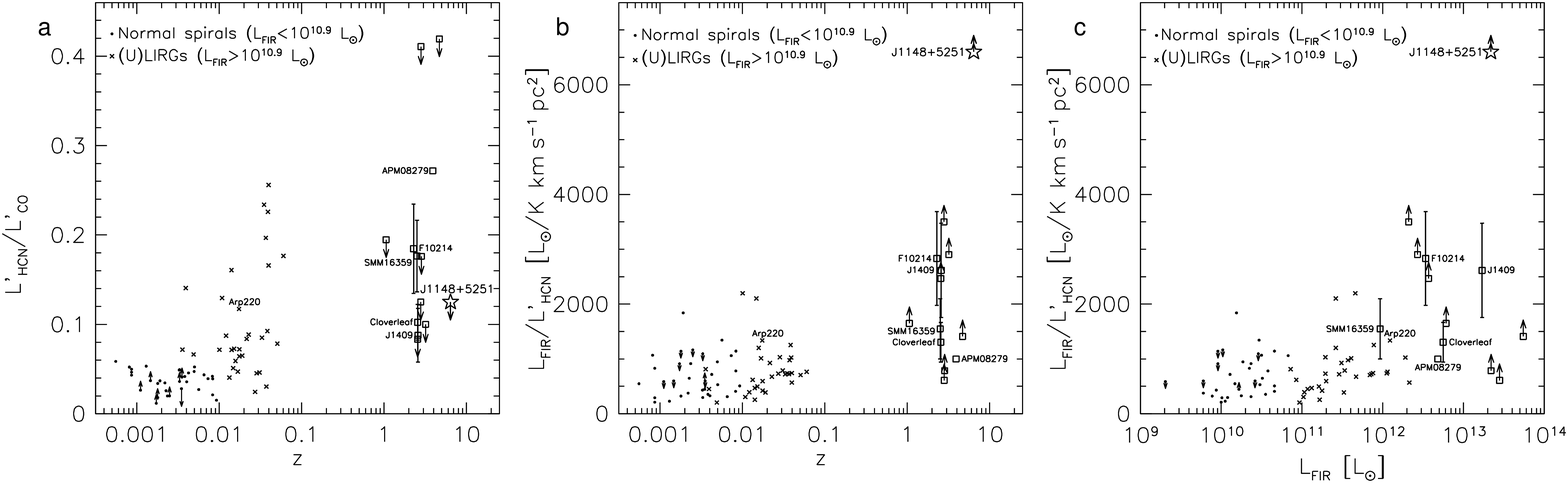}

\caption{Progression of the $L'_{\rm HCN}$/$L'_{\rm CO}$ (`dense gas
fraction') and $L_{\rm FIR}$/$L'_{\rm HCN}$ (`star formation law')
ratios with redshift and $L_{\rm FIR}$. Data for nearby spirals
($L_{\rm FIR}$$\lesssim$10$^{10.9}$\,\lsol\, i.e., $L_{\rm
  IR}$$<$10$^{11}$\,\lsol) and (U)LIRGs ($L_{\rm
  FIR}$$\gtrsim$10$^{10.9}$\,\lsol; GS04ab; G07), and high-$z$
galaxies including J1148+5251 are shown (detections labeled, see
Table~\ref{tab-2} for references). In the {\em left} panel,
down-arrows indicate upper limits for HCN, and up-arrows indicate that
HCN was detected, but not mapped throughout the whole galaxy. In the
{\em middle} and {\em right} panels, this is indicated vice versa.
\label{f2}}
\end{figure*}

The FIR luminosity is thought to originate dominantly from
dust-reprocessed light of young massive stars, and thus to be a good
tracer of the star-formation rate (SFR) in a galaxy. Star formation in
galaxies takes place in dense molecular clouds that are traced well by
HCN emission, resulting in a linear correlation between the FIR and
HCN luminosities (Wu \etal\ \citeyear{wu05}). However, G07 have found
a mild increase of $L_{\rm FIR}$/$L'_{\rm HCN}$ with increasing FIR
luminosity between normal spiral galaxies and ULIRGs [$L_{\rm
  FIR}$/$L'_{\rm HCN}$=580$^{+510}_{-270}$ (spirals)
vs.~740$^{+505}_{-50}$ (ULIRGs) on average; see Table~\ref{tab-2}].
This trend appears to get stronger toward the even more FIR-luminous
$z$$>$2 systems ($L_{\rm FIR}$/$L'_{\rm HCN}$=1525$^{+1300}_{-475}$ on
average). The lower limit of $L_{\rm FIR}$/$L'_{\rm HCN}$$>$6600
obtained for J1148+5251 is consistent with such a rising trend toward
high-$z$ systems, and extends it toward higher redshift
(Fig.~\ref{f2}b).  Due to the fact that the high-$z$ sources also have
systematically higher $L_{\rm FIR}$ than the nearby galaxies, it
remains unclear whether we observe a rising trend with redshift or
with $L_{\rm FIR}$ (Fig.~\ref{f2}c), or both.

\section{Discussion}

\subsection{Median Gas Density and Star Formation Efficiency}

Krumholz \& Thompson (\citeyear{kt07}) argue that $L_{\rm
  FIR}$/$L'_{\rm HCN}$ is expected to be higher for galaxies with a
median (molecular) gas density $n_{\rm med}$ close to or higher than
the critical density $n_{\rm crit}^{\rm HCN}$ required for excitation
of the observed HCN transition than for galaxies with lower $n_{\rm
  med}$.  In their case, they define star formation efficiency as the
fraction of the mass that is converted into stars per dynamical time
of the system. Note that this is different than the star formation
rate per unit total gas mass.  They argue that the non-linear relation
between $L_{\rm FIR}$ and $L'_{\rm CO}$ (e.g., Kennicutt
\citeyear{ken98a}, \citeyear{ken98b}; GS04b; Riechers \etal\
\citeyear{rie06b}) arises due to the fact that CO traces all gas.  The
star-formation rate is then dictated by the density $n$ divided by the
free-fall time $\tau_{\rm ff}$ ($\tau_{\rm ff} \propto n^{-0.5}$),
giving the standard Schmidt-law: star formation rate $\propto
n^{1.5}$, or $L_{\rm FIR}$ $\propto (L'_{\rm CO})$$^{1.5}$.  For
molecules like HCN, which only trace the small fraction of dense gas
clouds directly associated with star formation in normal galaxies,
$\tau_{\rm ff}$ is roughly fixed by $n_{\rm crit}$.  Hence the star
formation rate shows a linear relationship with $n$, or $L_{\rm FIR}$
$\propto (L'_{\rm HCN})$$^{1.0}$. However, in extreme galaxies, where
$n_{\rm med}$ in the molecular ISM approaches $n_{\rm crit}^{\rm
  HCN}$, $\tau_{\rm ff}$ again becomes relevant (i.e., HCN emission no
longer selects just the rare, dense peaks whose density is fixed by
$n_{\rm crit}^{\rm HCN}$, but instead traces the bulk of the ISM,
whose density can vary from galaxy to galaxy, and thus the variation
of $n$ and $\tau_{\rm ff}$ re-enter the calculation), and the
relationship approaches $L_{\rm FIR}$ $\propto (L'_{\rm HCN})$$^{1.5}$
(and $L'_{\rm HCN}$ $\propto L'_{\rm CO}$).  Interestingly, current
data show a marginal trend for a changing power-law index at the
highest luminosities of the type proposed by Krumholz \& Thompson.
This change in power-law index from 1 to 1.5 would suggest that, in
these extreme luminosity systems, $n_{\rm med}$ approaches $n_{\rm
  crit}^{\rm HCN}$.  More systems at high luminosity are required to
confirm this trend of changing power-law index.

\subsection{The Role of AGN Heating for $L_{\rm FIR}$}

Like most of the $z$$>$2 HCN-detected sources, J1148+5251 is a quasar.
It has been found that, even for such strong AGN galaxies, the bulk of
$L_{\rm FIR}$ is likely dominatly heated by star formation in most
cases (e.g., Carilli \etal\ \citeyear{car01}; Omont \etal\
\citeyear{omo01}; Beelen \etal\ \citeyear{bee06}; Riechers \etal\
\citeyear{rie06b}). However, based on radiative transfer models of the
dust SED of J1148+5251, Li \etal\ (\citeyear{li07}) argue that this
source may currently undergo a `quasar phase', in which AGN heating of
the hot and warm dust contributes significantly to $L_{\rm FIR}$. If
correct, this may be an alternative explanation for the elevated
$L_{\rm FIR}$/$L'_{\rm HCN}$ in this galaxy. The (rest-frame) IR
properties (tracing emission from hot dust) of J1148+5251 are similar
to those of other $z$$>$6 quasars with much lower $L_{\rm FIR}$
(tracing emission from warm dust), and even to local quasars (Jiang
\etal\ \citeyear{jia06}).  This supports the assumption that the hot
dust in J1148+5251 is dominantly heated by the AGN; however, the lack
of a correlation between $L_{\rm IR}$ and $L_{\rm FIR}$ in quasars
indicates that the warm dust may still be dominantly heated by star
formation.  Moreover, J1148+5251 follows the radio-FIR correlation for
star-forming galaxies (Carilli \etal\ \citeyear{car04}), which also
suggests a starburst origin for the dominant fraction of $L_{\rm
  FIR}$.

Furthermore, one of the $z$$>$2 HCN detections and some of the
meaningful limits are submillimeter galaxies without a known luminous
AGN, but are still offset from the local $L_{\rm FIR}$/$L'_{\rm HCN}$
relation.  It thus appears unlikely that AGN heating alone can account
for the higher average $L_{\rm FIR}$/$L'_{\rm HCN}$ in the high-$z$
galaxy sample.

\subsection{Implications for Future Studies}

Even when assuming the highest $L_{\rm FIR}$/$L'_{\rm HCN}$ of 2835
found among all HCN-detected galaxies in Table \ref{tab-2}, the depth
of our observations is sufficient to detect a galaxy with the redshift
and $L_{\rm FIR}$ of J1148+5251 ($z$=6.42) in HCN emission at a
signal-to-noise ratio of $>$4.5. To first order, our lower limit thus
is consistent with previous suggestions (G07) that $L_{\rm
  FIR}$/$L'_{\rm HCN}$ ratios in high redshift sources lie
systematically above those for nearby galaxies.  The scatter around
this trend is still significant, and will primarily be improved by
increasing the number of HCN-detected galaxies at high $z$.  In
addition, it will be important to improve on the main sources of error
for the individual high-$z$ detections (e.g., signal-to-noise limited
HCN/CO linewidth ratio, accuracy of the FIR SED fit, AGN bias of
$L_{\rm FIR}$).  The statistical and individual results, so far, would
even be consistent with an even stronger increase in $L_{\rm
  FIR}$/$L'_{\rm HCN}$ toward the highest $z$ and/or $L_{\rm FIR}$.
Our study of J1148+5251 may hint at such an effect.  Clearly, it is
desirable to obtain more sensitive observations of this source to
further investigate this issue.  Due to its superior collecting area
and high calibrational stability, the VLA is ideally suited for such a
sensitive study.  Although J1148+5251 is the most CO- and FIR-luminous
$z$$>$6 galaxy known, 80\,hr of VLA observations were necessary to
obtain the current limit.  In a favourable case, the \bhcn\ line may
have a strength of about 1.5 times the current rms. To obtain a solid
5\,$\sigma$ detection of such a line, of order 1000\,hr of
observations with the VLA would be required.  Due to improved
receivers and antenna performance, the fully operational EVLA will be
by a factor of two more sensitive to spectral lines of several
100\,\kms\ width (such as in J1148+5251), but will still require long
integration times. Studies of dense gas at $z$$>$6 thus appear to
require an order of magnitude increase in collecting area, such as
offered by future facilities like the SKA phase I demonstrator (e.g.,
Carilli \citeyear{car06}), which can serve as a low frequency
counterpart to the Atacama Large Millimeter/submillimeter Array
(ALMA).

\acknowledgments 
The authors would like to thank Philip Solomon and Yu Gao for access
to their dataset of nearby galaxies. We also thank Axel Wei\ss\ for
helpful discussions.  D.~R. thanks Vernesa Smol{\v c}i\'c for a great
piece of code.  D.~R.\ acknowledges support from the Deutsche
Forschungsgemeinschaft (DFG) Priority Program 1177.  C.~C.\
acknowledges support from the Max-Planck-Gesellschaft and the
Alexander von Humboldt-Stiftung through the Max-Planck-Forschungspreis
2005. We thank the referee for helpful comments.

\end{document}